\documentclass[conference, 10pt]{IEEEtran}

\IEEEoverridecommandlockouts

\usepackage{color}
\usepackage{bm}

\usepackage{color}
\usepackage{theorem}
\usepackage{times,amsmath,epsfig}
\usepackage{amssymb}
\usepackage{subfigure}
\usepackage{cite}
\usepackage{url}

\usepackage{algorithmic}
\usepackage{algorithm}

\usepackage{tikz}
\usetikzlibrary{shapes,arrows}
\input{mysymbol.sty}

\newcommand{\QED}{\hfill\ensuremath{\blacksquare}}

\title{A Robust Alternative for Graph Convolutional Neural Networks via Graph Neighborhood Filters}
\author{Victor M. Tenorio$^{*}$, Samuel Rey$^{*}$, Fer\hspace{0.015cm}nando Gama$^{\dag}$, Santiago Segarra$^{\dag}$, and Antonio G. Marques$^{*}$
	\thanks{Work in this paper was supported by the Spanish Grants SPGRAPH (PID2019-105032GB-I00), FPU17/04520, and CAM PEJ-2020-AI/TIC-18964, and the USA NSF award CCF-2008555. $^{*}$Dept. of Signal Theory and Comms., King Juan Carlos University, Madrid, Spain. $^{\dag}$ Dept. of ECE, Rice University, Houston, USA. Contact author: antonio.garcia.marques(AT)urjc.es.}}

\begin{document}
%
\maketitle
\begin{abstract}
Graph convolutional neural networks (GCNNs) are popular deep learning architectures that, upon replacing regular convolutions with graph filters (GFs), generalize CNNs to irregular domains. However, classical GFs are prone to numerical errors since they consist of high-order polynomials. This problem is aggravated when several filters are applied in cascade, limiting the practical depth of GCNNs.
To tackle this issue, we present the neighborhood graph filters (NGFs), a family of GFs that replaces the powers of the graph shift operator with $k$-hop neighborhood adjacency matrices.
NGFs help to alleviate the numerical issues of traditional GFs, allow for the design of deeper GCNNs, and enhance the robustness to errors in the topology of the graph.
To illustrate the advantage over traditional GFs in practical applications, we use NGFs in the design of deep neighborhood GCNNs to solve graph signal denoising and node classification problems over both synthetic and real-world data.
\end{abstract}
\begin{IEEEkeywords}
Graph Neural Networks, Graph Filters, Robust GSP, Non-Euclidean Data, Geometric Deep Learning.
\end{IEEEkeywords}
%

\section{Introduction}\label{S:Introduction}
The increasing complexity of current datasets, which oftentimes exhibit an underlying irregular structure, compels us to develop new models capable of learning efficiently from the observed data.
One alternative to exploit the irregular structure of the data at hand is provided by graph signal processing (GSP)~\cite{shuman2012theemergingfield,kolaczyk2014statistical,marques2020graph,ortega2018graph,djuric2018bookGSP}.
GSP is a rapidly growing field that assumes that the properties of the data are closely related to its underlying irregular structure, which can be accurately represented by a graph.
A prominent family of architectures provided by this discipline are graph convolutional neural networks (GCNNs)~\cite{Scarselli2009gnn, bronstein2017geometricdeeplearning, gama2020gnn}, which leverage the generalization of the convolution to irregular domains.
In recent years, these architectures have achieved state-of-the-art (SoA) performance in a wide range of applications involving graph-structured data~\cite{defferrard2016convolutional,kipf2017semisupervised,hamilton2017inductive,gama2019convolutionalgraphs,chowdhury2021unfolding,glaze2021principled,cutura2020deep}, including graph-signal denoising, which is used as an illustrative test case in the manuscript~\cite{chenSignalDenoising,rey2019underparametrized,onukiTrilateralDenoising}.

One of the key elements behind the success of GCNNs are graph filters (GFs)~\cite{SandryMouraSPG_TSP13,segarra2017filtering,bianchi2021graph}, which are linear operators that employ the structure of the graph to generalize the notion of classical convolution to graph signals.
To that end, GFs are defined as polynomials of the graph-shift operator (GSO), a matrix encoding the topology of the observed graph.
However, despite the success of GCNNs and the well-known benefits associated with classical GFs, the polynomial definition of the filters also comes with some limitations.
First, classical GFs are prone to numerical errors, a problem that is aggravated when several filters are placed in cascade~\cite{coutino2019advances}. This numerical instability can limit the design of deeper architectures.
Another relevant problem arises when there is uncertainty about the topology of the graph.
Because of their polynomial nature, GFs are sensitive to imperfections in the observed graph, harming the performance of the subsequent GSP tasks~\cite{segarra2016stability,ceci2020graph,natali2020topology,rey2021robust}.

Motivated by the previous discussion, we present a new type of linear graph-signal operators, referred to as neighborhood graph filters (NGFs), that replace the powers of the GSO with $k$-hop adjacency matrices.
These matrices encode the topological information of $k$-hop neighborhoods by capturing the existence of one or more shortest paths of a specific length between the nodes of a given graph. As a result, the output signal generated by an NGF can be interpreted as a linear combination of multiple signals, each of them consisting in the aggregation of the input values at the nodes located at a particular distance.  We discuss the main properties of the proposed NGFs with an especial focus on the numerical stability and the robustness to topology perturbations.
Furthermore, we exploit these filters to provide an alternative design of GCNN that is employed in the unsupervised task of graph signal denoising, and in the supervised problem of graph classification.
The performance of the resulting architecture is evaluated using both synthetic and real-world datasets.

The remainder of the paper is organized as follows. Section~\ref{S:fundamentals} provides basic concepts about GSP. 
Section~\ref{S:motivation} introduces the definition of NGFs and discusses their properties.
Section~\ref{S:arch} details the architecture resulting from combining GCNNs with NGFs, and Section~\ref{S:NumExSect} provides numerical validation of the presented architectures.
Concluding remarks in Section~\ref{S:conclusions} wrap up the paper.

\section{Fundamentals of graph signal processing}\label{S:fundamentals}

This section introduces notation and reviews basic GSP concepts that are leveraged throughout the paper.  

\vspace{.1cm}
\noindent\textbf{Graphs.} Let $\ccalG=\{\ccalV,\ccalE\}$ be a graph with $N$ nodes collected in the set $\ccalV$, and a set of edges $\ccalE$ such that $(i,j)$ belongs to $\ccalE$ if the nodes $i$ and $j$ are connected.
For any given $\ccalG$, its adjacency matrix is represented by the ${N\times N}$ matrix $\bbA$ with nonzero elements $[\bbA]_{ij}$ if and only if $(i,j)\in\ccalE$.
In other words, the adjacency matrix encodes the 1-hop neighborhoods of the graph. Note that for unweighted graphs, the entries of $\bbA$ are either $0$ or $1$. Finally, we use $\ccalN_i:=\{j: (i,j)\in\ccalE\}$ to denote the neighborhood of the node $i$ (i.e., the set of nodes that are linked to $i$); $d_{i,j}$ to denote the minimum number of hops between nodes $i$ and $j$; and $D=\max_{(i,j)}d_{i,j}$ to denote the diameter of $\ccalG$, which represents the length of the maximum shortest path present in the graph.

\vspace{.1cm}
\noindent\textbf{Graph signals.}  Signals observed on top of the graph are known as graph signals. More formally, a graph signal is a function $f: \ccalV\rightarrow \reals$ that can be represented as the vector $[x_1,...,x_N]^T\in\reals^N$, where $x_i=f(i)$ denotes the value of the signal observed at node $i$.
When modeling and processing graph signals, a key concept is the GSO $\bbS\in\reals^{N\times N}$~\cite{SandryMouraSPG_TSP13}, which is a matrix whose entries $(i,j)$ can be nonzero if and only if $(i,j) \in \ccalE$ or if $i=j$.
The matrix $\bbS$ captures the topology of the graph but makes no assumptions about the nonzero entries.
Typical choices for the GSO are the adjacency matrix $\bbA$~\cite{SandryMouraSPG_TSP13}, the graph Laplacian $\bbL = \diag(\bbA\mathbf{1}) - \bbA$~\cite{shuman2012theemergingfield}, and their respective generalizations.
The GSO $\bbS$ represents a linear transformation that can be computed locally at each node by aggregating the values of the input signal within the one-hop neighborhood of each of the nodes. 


\vspace{.1cm}
\noindent\textbf{GFs.} One of the most important tools in the context of GSP are GFs~\cite{SandryMouraSPG_TSP13,segarra2017filtering}.
GFs are linear graph signal operators that are defined as polynomials of the GSO. Mathematically, given an $N\times N$ matrix $\bbH$, the operator represented by the application of $\bbH$ to a graph signal is a GF if the matrix can be written as 
\begin{equation}\label{eq:classicgf}
    \bbH = \sum_{k=0}^{K-1} h_k \bbS^k,
\end{equation}
where $K-1$ denotes the degree of the filter and $\bbh=[h_0,...,h_{K-1}]$ collects the values of each of the $K$ filter coefficients. 
Since each application of $\bbS$ entails only exchanges among one-hop neighbors, when applying $\bbS^k$ to an input $\bbx$, the signal is being diffused across one-hop neighbors $k$ times. 
Effectively, this leads to a diffusion over a $k$-hop neighborhood~\cite{segarra2017filtering}. This readily implies that the output of a generic GF $\bbH$ to an input $\bbx$ can be written as $\bby=\bbH\bbx =\sum_{k=0}^K h_k (\bbS^k\bbx)=h_0 \bbx + h_1\bbS\bbx + h_2\bbS^2\bbx + ...$, i.e., a linear combination of graph signals, each of them corresponding to the original input diffused across neighborhoods of increasing size~\cite{segarra2017filtering}. 
The definition in \eqref{eq:classicgf} is easy to understand in the vertex domain, can be related to the classical definition of convolution and linear time-invariant (LTI) systems, has a neat spectral interpretation (see, e.g., \cite{SandryMouraSPG_TSP13} for the definition of the graph Fourier transform for signals and filters), and has been effectively used in a number of problems (such as denoising, deconvolution, or signal reconstruction, to name a few). However, one of its main limitations arises when $K$ is large, since high powers of $\bbS$ may render the filter (or the output) numerically ill-defined \cite{segarra2017filtering,coutino2019advances}. This issue is not present in classical LTI systems because the associated shift in the time domain is isometric, so that the energy (norm) of a signal is preserved regardless of the number of times the shift is applied. In the following sections, we explore some of these issues in further detail.


\section{Linear neighborhood graph filters}\label{S:motivation}
This section presents NGFs, a new type of linear operator for graph signals. 
The motivation for NGFs is to preserve most of the intuition present in classical polynomial GFs, while bypassing some of the numerical problems associated with high-degree filters. 

Let us set $\bbS=\bbA$ and suppose that $\ccalG$ is unweighted, so that the entries of $\bbA$ are binary. 
Then, it is well-known that $\bbA^k$ encodes the number of $k$-hop paths between any pair of nodes. 
For example, if $[\bbA^2]_{i,j}=3$, then, there are 3 paths of length 2 connecting nodes $i$ and $j$. 
Hence, for unweighted graphs, it follows that the application of $\bbA^k$ mixes the information within nodes that are at most $k$ hops away. 
It also demonstrates that as $k$ increases, the entries of $\bbA^k$ grow very large (the number of paths increases exponentially) leading to numerical issues. Our approach in this section is to replace $\bbA^k$ with a matrix that, while preserving the notion of $k$-neighborhood, does not grow arbitrarily large with $k$. 

To be precise, let us start by defining the $k$-hop adjacency matrices $\bbA_{\ccalN(k)}\in\{0,1\}$ as an $N\times N$ matrix whose entry 
$[\bbA_{\ccalN(k)}]_{ij}=1$ only if the nodes $i$ and $j$ are connected by at least one shortest path of exactly length $k$ (i.e., if the distance between $i$ and $j$ satisfies $d_{i,j}=k$).
Note that this definition implies $\bbA_{\ccalN(0)}=\bbI$ and $\bbA_{\ccalN(1)}=\bbA$.
With this notation at hand, we can then define the NGF $\bbH_\ccalN$ as a linear operator for graph signals that can be written as [cf. \eqref{eq:classicgf}]
\begin{equation}\label{eq:neighborgf}
    \bbH_{\ccalN} = \sum_{k=0}^{K-1}h_k\bbA_{\ccalN(k)},
\end{equation}
where $\bbh=[h_0,...,h_{K-1}]$ are the filter coefficients. As done in the previous section, when applied to an input graph signal $\bbx$, the NGF generates the output $\bby=\bbH_{\ccalN}\bbx$ where
\begin{equation}\label{eq:neighborgf_input_output}
    \bby= \sum_{k=0}^{K-1}h_k\bbA_{\ccalN(k)}\bbx=h_0\bbx + h_1\bbA_{\ccalN(1)}\bbx+ h_1\bbA_{\ccalN(2)}\bbx+...
\end{equation}
%

Unlike classical GFs, NGFs are not prone to numerical errors since the entries of $\bbA_{\ccalN(k)}$ are never larger than 1 and the norm $\|\bbA_{\ccalN(k)}\|$ need not increase with $k$.
As a result, NGFs are less sensitive to numerical instability. 
A particular interesting property of NGFs is that $\bbA_{\ccalN(k)}=\mathbf{0}_{N\times N}$ for all $k>D$, where $D$ represents the diameter of $\ccalG$, limiting the number of active filter coefficients to $D+1$. While this fact once again demonstrates the stability of NGFs, it also shows a potential loss on the expressiveness of the proposed filters, especially when $D$ is small.
This issue can be addressed by incorporating more expressive filters, e.g., node variant GFs~\cite{segarra2017filtering} or edge variant GFs~\cite{coutino2019advances}.
Nevertheless, the robustness to numerical issues derived from using the filters $\bbH_{\ccalN}$ as proposed in \eqref{eq:neighborgf} is expected to be more significant when the graph presents a high diameter $D$, as it is illustrated in Section~\ref{S:NumExSect}. 

Another relevant property of $\bbH_{\ccalN}$ is that, in computing each entry of the output, the value of the input signal at each node is considered at most once.
Indeed, if $[\bbA_{\ccalN(k')}]_{ij}=1$ then $[\bbA_{\ccalN(k)}]_{ij}=0$ for all $k \neq k'$.
Hence, assuming that $\ccalG$ is a connected graph and $K-1\geq D$, we have that
\begin{equation}\label{eq:add1}
    \sum_{k=0}^{K-1}\bbA_{\ccalN(k)}=\mathbf{1}_{N\times N},
\end{equation}
where $\mathbf{1}_{N\times N}$ denotes the $N \times N$ matrix of all ones.
Moreover, let $\barbA=\bbA+\bbE$ represent a perturbed adjacency matrix, with $\bbE$ denoting an error matrix that randomly removes or adds edges from $\bbA$, and let $\barbA_{\ccalN(k)}$ be the $k$-hop adjacency matrices of the perturbed graph associated with $\barbA$.
Since \eqref{eq:add1} holds for both $\bbA_{\ccalN(k)}$ and $\barbA_{\ccalN(k)}$, if the true and the perturbed graphs are connected graphs, when one link is removed several shortest paths are destroyed and the same number of shortest paths are created, and thus, for $K-1\geq D$ we have that
\begin{equation}\label{eq:0error_matrix}
    \sum_{k=0}^{K-1}\left(\barbA_{\ccalN(k)}-\bbA_{\ccalN(k)}\right)=\sum_{k=0}^{K-1}\bbE_{\ccalN(k)}=\mathbf{0}_{N\times N},
\end{equation}
where $\bbE_{\ccalN(k)}$ represents the error induced  by the matrix $\bbE$ in the $k$-hop adjacency matrix $\bbA_{\ccalN(k)}$.
From \eqref{eq:0error_matrix}, it can be observed that if all the filter coefficients are constant (i.e., $h_k=h$ for all $k$), the error between the perturbed and the true NGF is given by
\begin{equation}\label{eq:0error}
    \|\barbH_{\ccalN}-\bbH_{\ccalN}\|^2=\left\|h\sum_{k=0}^{K-1}\left(\barbA_{\ccalN(k)}-\bbA_{\ccalN(k)}\right)\right\|^2=0,
\end{equation}
where it can be seen that for this particular setting the NGF is impervious to topology perturbations.
GFs with constant coefficients appear, for example, in some message-passing applications~\cite{Zhang_2020_CVPR,dauwels2007variational}.

\begin{figure}
    \centering
    \includegraphics[width=0.35\textwidth]{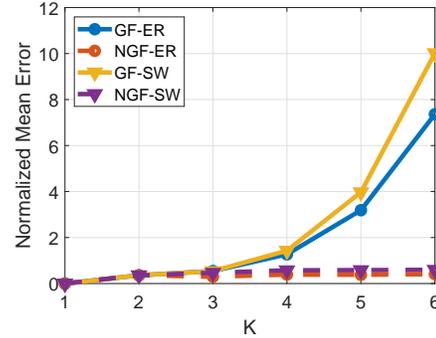}
    \vspace{-4mm}
    \caption{Evolution of the normalized error for classical GFs and NGFs as the number of filter taps $K$ increases for Erd\H{o}s-Rényi and small-world random graphs. The result is the mean error of 100 realizations.}
    \label{fig:err_filters}
\end{figure}

The preliminary result presented in \eqref{eq:0error} provides mathematical support to the idea that NGFs are more robust to topology imperfections than classical GFs.
The perturbed GF with $\barbA$ as the GSO is given by $\barbH=\sum_{k=0}^{K-1}h_k\barbA^k=\sum_{k=0}^{K-1}h_k(\bbA+\bbE)^k$, so intuitively, it can be observed that the powers of $\barbA$ will amplify the errors on $\bbA$, increasing the discrepancies between $\bbH$ and $\barbH$ as $k$ increases \cite{rey2021robust}.
On the other hand, regarding NGFs, when a path of length $k$ is perturbed due to imperfections in the observed adjacency matrix $\barbA$, the redundancy of paths existing in most graphs suggests that, as $k$ increases, it is more likely that an alternative path of the same length will also be available.
Therefore, it is expected that the difference between the matrices $\bbA_{\ccalN(k)}$ and $\barbA_{\ccalN(k)}$
decreases for higher values of $k$.
This intuition is evaluated numerically in Figure~\ref{fig:err_filters}, where the normalized errors  ${\|\barbH-\bbH\|_F^2}/{\|\bbH\|_F^2}$ and ${\|\barbH_{\ccalN}-\bbH_{\ccalN}\|_F^2}/{\|\bbH_{\ccalN}\|_F^2}$ are depicted for classical GFs and NGFs, respectively.
The comparison is carried out in two different random graphs: Erd\H{o}s Rényi and small-world graphs~\cite{newman2018networks}. In both cases, it can be seen that the error of classical GFs increases rapidly as $K$ grows.

This observation notwithstanding, the facts that matrices $\bbA_{\ccalN(k)}$ are not simultaneously diagonalizable and that they represent the presence of shortest paths (which is a non-differentiable operation) render the theoretical characterization of the robustness of $\bbH_{\ccalN}$ a challenging task that will be addressed in future works.


\section{NGF-based GCNN}\label{S:arch}
In this section, we introduce a natural extension of GCNNs, where the classical GFs are replaced by the novel NGFs. 

A graph neural network is a parametric non-linear function
\begin{equation}
f_{\bbTheta}(\bbZ|\ccalG):~\reals^{N\times F^{(0)}}\rightarrow\reals^{N\times F^{(L)}},    
\end{equation} 
that depends on the graph $\ccalG$.
The matrix $\bbZ$ represents the input of the architecture, the learnable weights are collected in $\bbTheta=\{\bbTheta^{(\ell)}\}_{\ell=1}^L$, $L$ represents the number of layers, and $F^{(\ell)}$ denotes the number of features at layer $\ell$.
When designing a graph neural network there are several alternatives to account for the topology of the graph.
One common approach used in specific implementations of GCNNs exploits the message passing operation.
The resulting architecture is given by the following recursion
\begin{equation}\label{eq:basicgnn}
    \bbX^{(\ell)} = \sigma \big( \bbA \bbX^{(\ell-1)} \bbTheta^{(\ell)} \big),
\end{equation}
where $\sigma(\cdot)$ is an element-wise non-linear function typically known as activation function, the weights $\bbTheta^{(\ell)}$ are matrices of dimension $F^{(\ell-1)}\times F^{(\ell)}$, and $\bbX^{(\ell)}\in\reals^{N \times F^{(\ell)}}$ denotes the output of the $\ell$-th layer with $\bbX^{(0)}=\bbZ$ and $\bbX^{(L)}$ being the input and the output of the architecture, respectively.
Note that we can interpret the computation of $\bbX^{(\ell)}$ as first left-multiplying the input of the previous layer by the adjacency matrix of the graph $\bbA$, which combines the signal values of neighboring nodes, and then right-multiplying by the learnable weights $\bbTheta^{(\ell)}$, so the architecture learns to combine different features from previous layers.

The main disadvantage of the architecture presented in \eqref{eq:basicgnn} is that, at each layer, the convolution is carried out only in the 1-hop neighborhoods.
This effect is mitigated by stacking $L$ layers, since it is roughly equivalent to perform a convolution on the $L$-hop neighborhood.
However, this alternative produces an undesirable coupling between the depth of the GCNN and the size of the neighborhood where the convolution is applied.

\begin{figure*}[!t]
\centering
\centerline{
\subfigure[]{\includegraphics[width=0.5\textwidth]{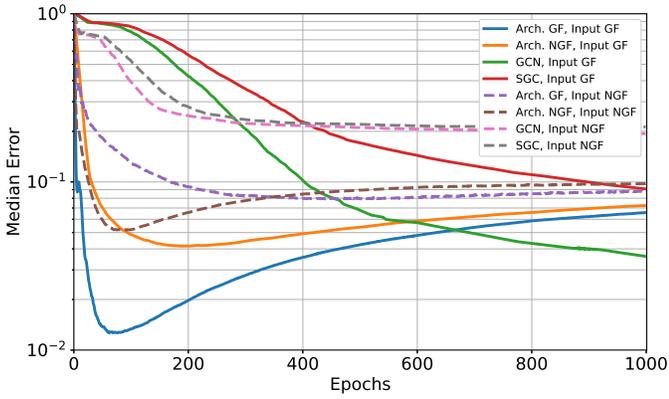} \vspace{-.3cm}\label{sfig:denoising_2l}}
\subfigure[]{\includegraphics[width=0.5\textwidth]{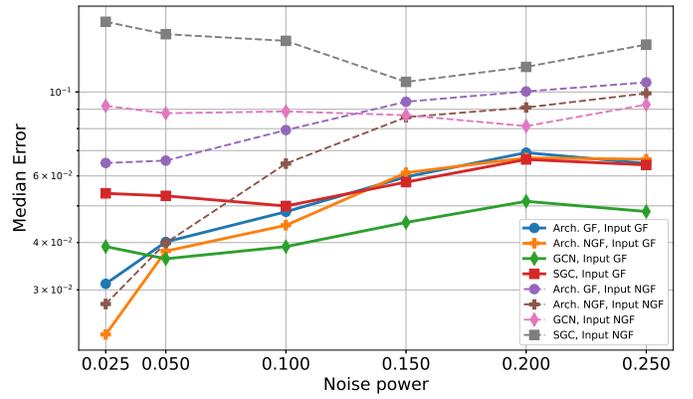} \vspace{-.3cm}\label{sfig:denoising_dd}}
\vspace{-.2cm}}
\caption{Comparison of the performance of the proposed NGF against a classical GF setting in the synthetic graph signal denoising task. (a) Median error as a function of training epochs for a bandlimited signal corrupted by noise with power $0.1$. (b) Median error achieved by the architectures as a function of noise power when denoising a diffused white signal (for each experiment, the number of epochs corresponds to the one that minimizes the error). In both cases, the signal is normalized to have unit norm. \vspace{-.2cm}}
\label{fig:results_denoising}
\end{figure*}

One way to avoid the aforementioned coupling is to replace the matrix $\bbA$ with a graph filter $\bbH$, resulting in an architecture implemented by the recursion 
\begin{equation}\label{eq:gcnn}
    \bbX^{(\ell)} = \sigma \big( \bbH^{(\ell)} \bbX^{(\ell-1)} \bbTheta^{(\ell)} \big).
\end{equation}
Note that $\bbH^{(\ell)}$ is a classical filter of the form \eqref{eq:classicgf} and the filter coefficients can be: 1) fixed in advance as a prior or 2) learned by the architecture, giving rise to two different architectures. Although GFs decouple the depth of the architecture from the range of the convolution, the drawbacks introduced in Section~\ref{S:motivation} still impose some limitations on the architecture.
One evident restriction affects the depth of the architecture.
Stacking layers is similar to applying several GFs in cascade, so the numerical issues derived from high-order polynomials can give rise to exploding or vanishing gradients (especially when ReLUs are used as activation functions).
To circumvent this problem, in this work we replace the classical GF with the NGF, so the neighborhood GCNN (NGCNN) is given by
\begin{equation}
        \bbX^{(\ell)} = \sigma \big( \bbH_{\ccalN}^{(\ell)} \bbX^{(\ell-1)} \bbTheta^{(\ell)} \big),
\end{equation}
where $\bbH_{\ccalN}^{(\ell)}$ is a filter of the form \eqref{eq:neighborgf} and the filter coefficients can be either fixed in advance or learned by the architecture.
The proposed NGCNN preserves the structure of the graph-aware per-layer linear transformation, so the weights are only learning to mix the different features of the input matrices while the relation between the signal values at the different nodes is determined by the NGF.
Note that, from a GSP perspective, since the matrix $\bbX^{(\ell-1)}$ is interpreted as $F^{(\ell-1)}$ different graph signals, the product $\bbH_{\ccalN}^{(\ell)} \bbX^{(\ell-1)}$ is seen as a node-domain convolution.
Moreover, thanks to the filters $\bbH_{\ccalN}^{(\ell)}$, the depth of the architecture and the range of the convolutions are completely decoupled since the NGF are less prone to numerical issues, and thus, they do not impose a limiting factor in the depth of the architecture. On the other hand, the maximum degree of the NGF at each layer is $D$, the diameter of the graph. 

Next, we evaluate the performance of the proposed NGCNN in different settings and numerically illustrate how NGFs mitigate some of the typical limitations of GCNN.

\section{Numerical Simulations}\label{S:NumExSect}

We analyze the performance of the proposed NGF in two different graph-related problems: signal denoising using synthetic data (Section~\ref{Sub:Denoising}) and node classification in three real-world citation networks (Section~\ref{Sub:nodeClassification}).
The python code used to run the experiments, which describes in detail all the settings of the proposed architecture, is available online\footnote{\url{https://github.com/vmtenorio/NeighborhoodGF}}.

\subsection{Graph signal denoising}\label{Sub:Denoising}

The goal in graph-signal denoising is to recover the original graph signal $\bbx\in\reals^{N}$ given the noisy graph-signal observation $\bby=\bbx+\bbw$, with $\bbw\in\reals^{N}$ representing a noise vector.
To that end, we approach the denoising problem as in~\cite{rey2019underparametrized,do2020graph} by minimizing
\begin{alignat}{2}\label{signal_denoising1}
\!\!&\!\hbTheta = \argmin_{\bbTheta} && \|\bby-\bbf_{\bbTheta}(\bbZ|\ccalG)\|_2^2,
\end{alignat}
where the entries of the input matrix $\bbZ$ are randomly sampled from a zero-mean unit-variance normal distribution.
The problem~\eqref{signal_denoising1} is minimized by running gradient descent for a fixed number of epochs.
After estimating the weights $\hbTheta$, the denoised signal is given by $\hbx=\bbf_{\hbTheta}(\bbZ|\ccalG)$.
The motivation behind this approach is that the proposed architecture is capable of learning the signal $\bbx$ faster than the noise $\bbw$ and, thus, early stopping can be applied to learn most of the signal without learning too much noise.

To analyze the performance of the proposed filters, we created synthetic graph signals defined over a random stochastic-block-model (SBM)~\cite{newman2018networks} graph.
In this graph-generative model, nodes are divided into communities and edges are randomly drawn between nodes of the same community independently with a probability of 0.3, and between nodes of different communities, also independently, with a probability of 0.0075.
The graph contains 256 nodes and 8 or 4 equally-sized communities for test cases 1 and 2, respectively.

\begin{figure*}[!t]
\centering
\centerline{
\subfigure[]{\includegraphics[width=0.5\textwidth]{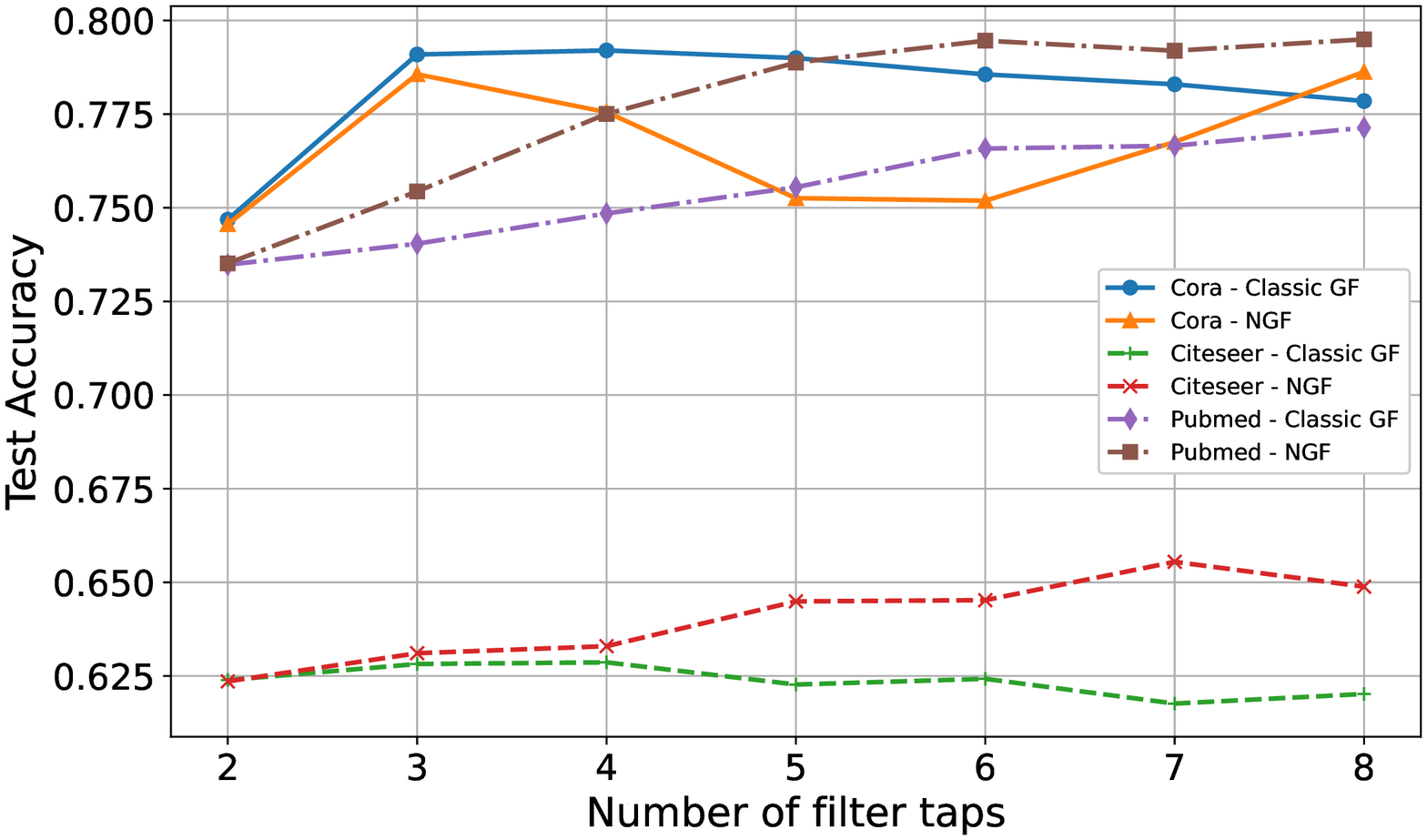}\vspace{-.3cm} \label{sfig:kanalysis}}
\subfigure[]{\includegraphics[width=0.5\textwidth]{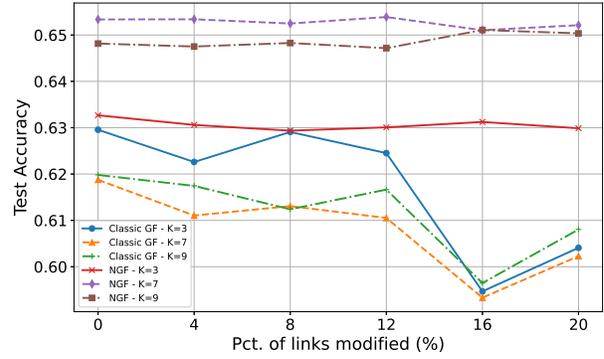} \vspace{-.3cm}\label{sfig:kpert}}
\vspace{-.2cm}}
\caption{Comparison of the performance of the proposed NGF against a classical GF setting in three real-world scenarios.
(a)~Test classification accuracy for the three considered datasets as a function of the number of filter coefficients used in the network. 
(b)~Performance of the architectures using the Citeseer dataset, for different numbers of filter taps ($K$) as a function of the amount of perturbations introduced to the graph.\vspace{-.3cm}}
\label{fig:results_real}
\end{figure*}

\vspace{1mm}\noindent
\textbf{Test case 1.} The results for the first test case are shown in Figure~\ref{sfig:denoising_2l}.
The signal $\bbx$ is created by diffusing a zero-mean white signal $\bbb\in \reals^N$ over the graph as $\bbx=\bbH\bbb$ for the case ``Input GF'', and $\bbx=\bbH_{\ccalN}\bbb$ for ``Input NGF''.
Both filters $\bbH$ and $\bbH_{\ccalN}$ have random coefficients $\bbh$, drawn from a white uniform distribution between 0 and 1 and are normalized so that $\bbh^T \mathbf{1} = 1$.
We add zero-mean white Gaussian noise $\bbw$ with normalized power of $0.1$, and generate 200 different signals $\bbx$.
The results shown in the figure report the median error $\|\bbx-\hbx\|_2^2/\|\bbx\|_2^2$ across the 200 realizations as the number of epochs increases for four different 2-layer architectures: one using the classical GF (``Arch. GF''), another one using the NGF (``Arch. NGF''), and two SoA architectures: Graph Convolutional Networks~\cite{kipf2017semisupervised} (GCN) and Simple Graph Convolution~\cite{wu19sgc} (SGC).
It can be observed that the error starts decreasing, reaches a minimum, and then increases.
This latter increase is due to the architecture starting to fit the noise, so early stopping is needed to obtain the denoised $\hbx$.
It can also be seen that the performance is tightly coupled to the signal generation method.
If we generate $\bbx$ using $\bbH$ then the architecture with the best performance is the one using a classical GF, beating both SoA architectures in terms of minimum error.
However, when generating the signal with the NGF, the architecture with superior performance and the only one able to denoise the signal is the one using an NGF.
As a result, NGCNN is the only architecture to effectively denoise the signals in the two scenarios considered.


\vspace{.5mm}\noindent
\textbf{Test case 2.}
The signal $\bbx$ is generated as in the previous test case.
The results are shown in Figure~\ref{sfig:denoising_dd}, where we depict the evolution of the normalized error achieved by each architecture as the normalized power of the noise increases.
The figure reveals that NGFs and classic GFs behave similarly when the input is generated using a classic GF, with GCN outperforming both architectures for high noise power values.
However, NGF clearly outperforms classic GF in all cases when the signal is generated with a diffusion process that utilizes an NGF, illustrating that NGFs are more flexible linear operators. NGCNN also outperforms GCN in low noise power settings, while GCN achieves a lower minimum error by a small margin when the noise power is greater than 0.2

\subsection{Node Classification}\label{Sub:nodeClassification}

We analyzed the performance of the proposed graph filters using 3 real-world datasets, where the graph represents a network of publications (nodes are published articles and edges represent citations between them) and the goal is to classify each node into a category that denotes the topic of the publication.
The graph signals in each node indicate the presence or absence of words from a dictionary.
Further details of these datasets can be found in Table~\ref{tab:datasets} and \cite{mccallum_2000_cora,giles_1998_citeseer,prithviraj_2008_pubmed}.

To perform the classification task, the output of the architectures is given by $\bbX^{(L)} \in \reals^{N \times M}$, where $M$ denotes the total number of classes.
The non-linearity in the last layer is given by the softmax function so that the features in each node are interpreted as a measurement of the probability for the specific node to belong to a certain class.
The architectures are trained with the cross-entropy loss, a typical loss function used for classification problems.
To avoid numerical issues related to the powers of the GSO, in these experiments we normalize the GSO by its largest eigenvalue $\tbS = \bbS / \lambda_{max}$.

\begin{table}[ht]
    \centering
    \caption{Datasets used for node classification.}
    \vspace{-.1cm}
    \begin{tabular}{c|c|c|c|c|c}
        Dataset name & \#Nodes & \#Features & \#Classes & Radius* & Diameter* \\ \hline
        Cora~\cite{mccallum_2000_cora} & 2708 & 1433 & 7 & 10 & 19 \\
        Citeseer~\cite{giles_1998_citeseer} & 3327 & 3703 & 6 & 15 & 28 \\
        Pubmed~\cite{prithviraj_2008_pubmed} & 19717 & 500 & 3 & 10 & 18
    \end{tabular}
    \newline \\
    * Metrics refer to the largest connected component of the graph.
    \label{tab:datasets}
\end{table}

\vspace{1mm}\noindent
\textbf{Test case 3.}
With the configuration described above, the accuracy over the set of test nodes obtained in each dataset for both types of GFs can be seen in Figure~\ref{sfig:kanalysis}.
The image shows the performance of the GCNN and the NGCNN as the order of the filters increases.
It can be seen that, for $K=2$, the performance of both filters is similar.
This is not surprising since filters of this order only take into account the 1-hop neighborhood, so in this case $\bbH=\bbH_{\ccalN}$.
Nevertheless, as $K$ increases the behavior aligns with the discussion provided in Section~\ref{S:motivation}.
The performance of the NGF improves while the performance of the classical GF deteriorates or improves at a much slower rate.
This illustrates how NGFs are more robust to numerical issues related to higher-order filters
This can be seen especially clear for the Citeseer dataset, which has the graph with the highest diameter among the three.

\vspace{1mm}\noindent
\textbf{Test case 4.}
In this case the focus is on the Citeseer dataset, and we analyze the performance of the network as we increase the perturbation introduced in the graph; see Figure~\ref{sfig:kpert}.
The effect of the perturbation is measured for different values of the number of filter coefficients $K$, as indicated in the legend.
The perturbations consist in randomly removing and creating links in the original graph, as a percentage of the total number of existing links.
NGFs outperform classic GF in all the tested settings and the performance of NGFs remains approximately constant independently of the perturbation introduced, while the accuracy of the architectures with classical GFs decreases.
This result is aligned with the discussion presented throughout this paper, showcasing that NGFs are more robust to errors in the topology of the given graph and further motivating the development of associated theoretical results.

\section{Conclusions}\label{S:conclusions}
This paper proposed graph neural network architectures for graph signals based on NGFs, a new type of GFs whose definition leverages the (adjacency) matrices encoding the $k$-hop neighborhood of the nodes of the graph. Compared with classical GFs, the proposed NGFs architectures are i)~more stable when the number of filter coefficients increases and ii)~more robust against graph perturbations, rendering this new technique more suitable for problems where the graph is not known with certainty.
We validated these claims through experimental results in synthetic datasets and, more conspicuously, in real-world datasets with larger graphs in terms of both size and diameter. Future work includes the theoretical characterization of the robustness of NGFs as well as additional simulations in real-world tasks.


\bibliographystyle{IEEEbib}
\bibliography{myIEEEabrv,biblio}

\end{document}